\documentclass[sigconf,nonacm]{acmart}
\acmConference[WSDM '26]{ACM International Conference on Web Search and Data Mining}{Feb 22--26, 2026}{Boise, USA}
\acmYear{2026}\copyrightyear{2026}

\usepackage{enumitem} 
\usepackage{multirow}
\usepackage{hyperref}
\usepackage{booktabs}  
\usepackage{makecell}  

\newcommand{\Method}{\textsc{CoRank}}

\newcommand\para[1]{\smallskip\noindent\textbf{#1}}

\begin{document}

\title{CoRank: LLM-Based Compact Reranking with Document Features for Scientific Retrieval}

\author{Runchu Tian}
\affiliation{%
  \institution{University of Illinois \\Urbana-Champaign}
  \city{Urbana}
  \country{United States}}
\email{runchut2@illinois.edu}

\author{Xueqiang Xu}
\affiliation{%
  \institution{University of Illinois \\Urbana-Champaign}
  \city{Urbana}
  \country{United States}}
\email{xx19@illinois.edu}

\author{Bowen Jin}
\affiliation{%
  \institution{University of Illinois \\Urbana-Champaign}
  \city{Urbana}
  \country{United States}}
\email{bowenj4@illinois.edu}

\author{SeongKu Kang}
\affiliation{%
  \institution{Korea University}
  \city{Seoul}
  \country{South Korea}}
\email{seongku@korea.ac.kr}

\author{Jiawei Han}
\affiliation{%
  \institution{University of Illinois \\Urbana-Champaign}
  \city{Urbana}
  \country{United States}}
\email{hanj@illinois.edu}

\begin{abstract}
Scientific retrieval is essential for advancing scientific knowledge discovery. 
Within this process, document reranking plays a critical role in refining first-stage retrieval results.
However, standard LLM listwise reranking faces challenges in the scientific domain.
First-stage retrieval is often suboptimal in the scientific domain, so relevant documents are ranked lower. Meanwhile, conventional listwise reranking places the full text of candidates into the context window, limiting the number of candidates that can be considered.  As a result, many relevant documents are excluded before reranking, constraining overall retrieval performance.
To address these challenges, we explore semantic-feature-based compact document representations (e.g., categories, sections, and keywords) and propose \Method, a \textit{training-free}, \textit{model-agnostic} reranking framework for scientific retrieval.
It presents a three-stage solution: (i) offline extraction of document features, (ii) coarse-grained reranking using these compact representations, and (iii) fine-grained reranking on full texts of the top candidates from (ii). This integrated process addresses suboptimal first-stage retrieval: Compact representations allow more documents to fit within the context window, improving candidate set coverage, while the final fine-grained ranking ensures a more accurate ordering.
Experiments on 5 academic retrieval datasets show that \Method~significantly improves reranking performance across different LLM backbones (average nDCG@10 from $50.6$ to $55.5$). 
Overall, these results underscore the synergistic interaction between information extraction and information retrieval, demonstrating how structured semantic features can enhance reranking in the scientific domain.
\end{abstract}



\maketitle


\section{Introduction}
\begin{figure}[t]
  \centering
  \includegraphics[width=0.7\linewidth]{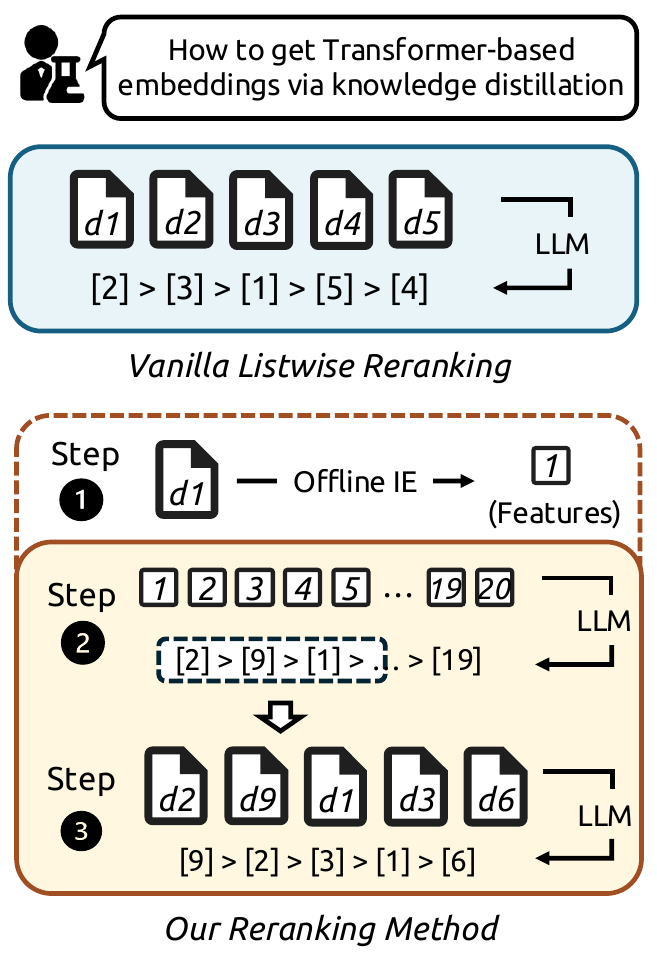}
  \caption{We extract semantic features, rerank a larger candidate pool with them, and finally refine the shortlist with full text instead of directly reranking with full text.}
  \Description{Flow diagram with three stages: (i) offline extraction of document-level features; 
  (ii) coarse-grained reranking over a large candidate set using compact feature vectors; 
  (iii) fine-grained reranking of the shortlisted candidates using full-text representation. 
  Arrows indicate the processing order from (i) to (iii).}
  \label{fig:sec_1_profile}
  \vspace{-1.5em}
\end{figure}

Scientific retrieval~\citep{lawrence1999indexing, white2009scientific} is crucial for scientific knowledge discovery. While current retrievers are effective at retrieving broadly relevant scientific papers, they often struggle to differentiate among documents covering similar topics~\citep{Sciavolino2021SimpleEQA, Liu2021DenseHRA}, failing to make fine-grained relevance estimation. Therefore, the reranking stage~\citep{Carbonell1998TheUOA, Kurland2005PageRankWHA} is particularly important~\citep{Gao2021RethinkTOA}, as it refines the first-stage retrieval to better distinguish between closely ranked documents.

Recently, Large Language Models (LLMs)~\citep{grattafiori2024llama, openai2024gpt4o} have significantly advanced the field of document reranking, particularly through their application in \textit{listwise reranking}~\citep{sun2023chatgpt, ma2023zeroshotlistwisedocumentreranking}. In this setting, LLMs jointly assess a set of retrieved candidates within their context window and generate a reordered ranking based on their relevance.
Previous works~\citep{pradeep2023rankvicunazeroshotlistwisedocument, gangi-reddy-etal-2024-first, liu2024slidingwindowsendexploring,ren2024selfcalibratedlistwisererankinglarge} show that LLM-based listwise rerankers outperform prior embedding-based approaches~\citep{nogueira2019multistagedocumentrankingbert, nogueira-etal-2020-document} on multiple evaluation datasets~\citep{thakur2021beirheterogenousbenchmarkzeroshot}.

The standard practice in LLM listwise reranking~\citep{sun2023chatgpt, pradeep2023rankvicunazeroshotlistwisedocument, pradeep2023rankzephyreffectiverobustzeroshot, gangi-reddy-etal-2024-first, liu2024slidingwindowsendexploring} is to input the full text of each candidate document into the context window. In scientific retrieval, however, this approach faces key limitations. In the scientific domain, first-stage retrievers often show limited performance~\citep{Kim2023RelevanceassistedGFA, Kang2024ImprovingRI, kang2024taxonomy}, so truly relevant documents may not rank high enough in the first-stage retrieval. Moreover, due to the substantial token overhead of full text representation and finite context length, rerankers can only operate on a limited number (typically 20 per prompt)~\citep{sun2023chatgpt, ma2023zeroshotlistwisedocumentreranking} of candidates. Consequently, when the first-stage retrieval is suboptimal, many relevant documents are excluded from reranking consideration, which inherently constrains overall performance~\citep{reddy2024refitrelevancefeedbackreranker}.


A straightforward solution is to expand the context window of LLM rerankers to handle more candidates. This can be done using long-context LMs~\citep{huang2023advancing, pawar2024and} or test-time scaling strategies~\citep{snell2024scaling, xia2025generativeaiactii}, such as multi-round reranking~\citep{sun2023chatgpt, ma2023zeroshotlistwisedocumentreranking}. However, these methods have notable drawbacks: long-context LMs require additional training~\citep{liu2024slidingwindowsendexploring}, suffer from “loss in the middle” problem~\citep{liu2023lostmiddlelanguagemodels, Hsieh2024RULERWTA}, and incur high computational costs~\citep{liu2025survey}; sliding windows similarly increase inference latency and overhead~\citep{sun2023chatgpt, ma2023zeroshotlistwisedocumentreranking}. To better address the limitations, we explore compressing each document’s representation by abstracting high-level features (e.g., categories, keywords), enabling more documents to fit within a fixed context window. This achieves a similar effect to window expansion but with greater efficiency.

We propose \Method, a \textit{zero-shot}, \textit{model-agnostic} reranking framework for science retrieval as exhibited in Figure~\ref{fig:sec_1_profile}. \Method \ consists of three stages:
(i) \textbf{Offline Preprocessing}: extract high-level semantic features like categories and keywords from unstructured scientific documents; 
(ii) \textbf{Coarse-grained Reranking}: use these compact representations for an initial ranking and select a subset of top candidates; 
(iii) \textbf{Fine-grained Reranking}: rerank \textit{the top candidates} using full scientific documents to recover the potentially missing details during information extraction. This integrated reranking framework aims to address the suboptimal first-stage retrieval. Compact representations allow more documents to fit within the context window, improving candidate set coverage, while the final fine-grained ranking ensures a more accurate ordering.

We comprehensively evaluate \Method~on five diverse scientific retrieval datasets spanning computer science~\citep{ajith2024litsearchretrievalbenchmarkscientific, mysore2021csfcubetestcollection}, biomedical~\citep{boteva2016full}, and clinical~\citep{wadden-etal-2020-fact, trec-covid} domains. Our experiments cover a diversity of LLM backbones, including both proprietary~\citep{openai2024gpt4o, google2025gemini} and open-source~\citep{qwen2025qwen25technicalreport, jiang2023mistral7b, grattafiori2024llama} models. Results show that \Method~consistently outperforms existing LLM-based zero-shot reranking baselines, achieving an average absolute improvement of $+4.9$ nDCG@10. Notably, our method delivers strong gains even in challenging settings with suboptimal first-stage retrieval, while also maintaining high efficiency by reducing token consumption compared to the widely adopted sliding window strategy~\citep{sun2023chatgpt}. Extensive ablation studies and further analyses validate the contribution of each component and demonstrate the robustness of our method.

\begin{enumerate}[label=\textbf{\# \arabic*}, leftmargin=*, itemsep=4pt, parsep=0pt, topsep=4pt, partopsep=0pt]
\item We reveal the limitations of LLM listwise reranking in scientific retrieval.
\item We explore semantic features for document representation in reranking and propose \Method, a novel zero-shot reranking framework.
\item We show that \Method \ significantly improves reranking performance of LLM-based listwise approach in scientific retrieval.
\end{enumerate}

These findings underscore how semantic information extraction and retrieval interact, with pre-analysis of document features enabling more effective reranking in scientific search.

\section{Related Work}

\subsection{Classical Document Reranking}

Document reranking~\citep{karpukhin2020dense, Ren2021PAIRLP, Ren2023TOMEAT} is a critical component in information retrieval (IR)~~\citep{baeza1999modern, singhal2001modern}, especially when high retrieval precision is required and the first-stage retriever alone is insufficient. Early reranking methods were lexical and probabilistic~~\citep{Salton1975AVS, Robertson1976RelevanceWO}. Like early sparse retrieval methods~\citep{baeza1999modern, robertson2009probabilistic}, they relied on term overlap between the query and document to adjust relevance scores. However, unlike initial retrieval, rerankers \textit{revisit} a smaller set of candidate documents returned by the first stage and apply more sophisticated weighting or adjustment techniques to refine the ranking. Nevertheless, these approaches were limited by their reliance on exact matching and failed to capture in-depth semantics~\citep{karpukhin2020dense, Gao2021COILREA}.

With advances in deep learning, neural reranking methods~\citep{Guo2019ADL, Trabelsi2021NeuralRM}, particularly those based on learning-to-rank~\citep{Cao2007LearningTR, Liu2009LearningTR} frameworks, became prevalent. These models overcome previous limitations, capturing semantic similarity beyond surface-level term matches. The emergence of pre-trained language models (PLMs)~\citep{Vaswani2017AttentionIA, Devlin2019BERTPO, Reimers2019SentenceBERTSE} further transformed document reranking~\citep{Nogueira2019FromDT, Lin2020PretrainedTF}. Cross-encoder-based rerankers~\citep{Litschko2022ParameterEfficientNR}, in particular, have demonstrated strong performance by modeling query-document relevance scores and optimizing pointwise, pairwise, or listwise loss~\citep{Zhuang2022RankT5FT}. Unlike bi-encoder architectures used in dense retrieval~\citep{karpukhin2020dense}, cross-encoders jointly encode both the query and each candidate document, allowing the model to directly capture rich dependencies between queries and documents from first-stage retrieval. 

\begin{figure*}[t]
  \centering
  \includegraphics[width=0.97\linewidth]{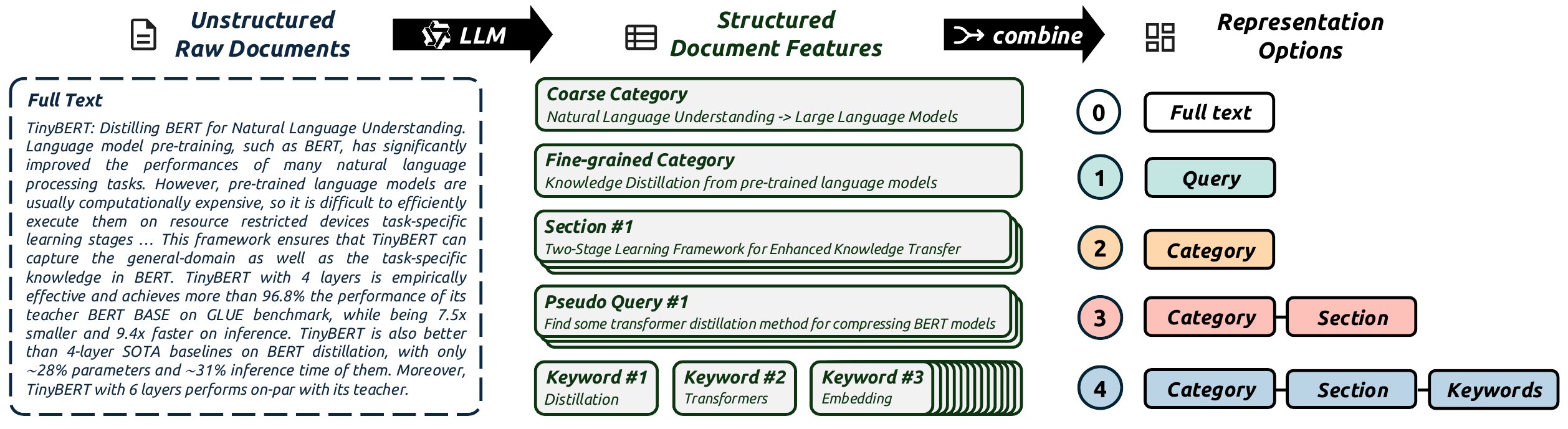}
  \caption{Overview of our feature extraction pipeline: from unstructured documents, we apply zero-shot LLM information extraction to obtain document features including categories, sections, pseudo queries, and keywords, which are then combined into compact representations.}
  \Description{Flow diagram of the document representation pipeline. Left: an unstructured scientific document enters the system. Middle: a zero-shot LLM information extraction step produces four feature sets—categories (three-level path), section titles, pseudo queries, and keywords. An adaptive selection module keeps the most relevant items (e.g., one section, one pseudo query, and top-k keywords). Right: the selected items are concatenated into a compact representation that is passed to the reranker. Arrows indicate the order: document → extraction → selection → compact representation → reranking.}
  \label{fig:representation_pipe}
  \vspace{-1em}
\end{figure*}

\subsection{LLM-Based Listwise Document Reranking}

Recently, large language models (LLMs)~\citep{touvron2023llama2openfoundation, openai2024gpt4technicalreport} have revolutionized the field of NLP. With strong general knowledge and instruction-following~\citep{Ouyang2022TrainingLM} capabilities, LLMs offer a new paradigm for listwise document reranking. A classic proposal~\citep{sun2023chatgpt} involves placing the query and a list of candidate documents into the context window and prompting the LLM to generate a reordered ranking. This zero-shot strategy has been shown to outperform the traditional reranking baselines~\citep{sun2023chatgpt, ma2023zeroshotlistwisedocumentreranking}.

After the introduction of LLM-based listwise reranking, research has rapidly extended in several directions. For \textit{model development}, works such as RankVicuna~\citep{pradeep2023rankvicunazeroshotlistwisedocument} and RankZephyr~\citep{pradeep2023rankzephyreffectiverobustzeroshot} pioneered open-source reranking LLMs, while RankMistral~\citep{liu2024slidingwindowsendexploring} leveraged long-context training to adapt LLMs for larger document lists in a single input. For \textit{context-length adaptation}, Sliding window strategies~\citep{sun2023chatgpt, ma2023zeroshotlistwisedocumentreranking} partition candidate lists into overlapping chunks for broader coverage, addressing the limited context windows of LLMs. For \textit{inference efficiency}, FIRST~\citep{gangi-reddy-etal-2024-first} improves the speed of listwise reranking by restricting the process to first-token decoding. \citet{Liu2024LeveragingPEA} and \citet{Li2024KeyB2SKA} use passage embeddings and chunked text blocks to represent documents reducing the token cost per document. From the perspective of \textit{self-consistency}, ScaLR~\citep{ren2024selfcalibratedlistwisererankinglarge} introduces self-calibration to improve consistency across windows. Meanwhile, \citet{Tang2023FoundIT} enhance reranking quality through permutation consistency.

Since our method operates on document representations in a \textit{plug-and-play} fashion, it is orthogonal to most previous baselines and can be applied jointly with them (we illustrate this with the sliding window strategy in Section~\ref{sec: experiments}). Notably, two prior studies~\citep{Liu2024LeveragingPEA, Li2024KeyB2SKA} have also identified limitations of full-document representations, but their approaches are fundamentally different from ours: one~\citep{Liu2024LeveragingPEA} adopts non-natural language embeddings, while the other~\citep{Li2024KeyB2SKA} merely relies on document chunking. Furthermore, their analyses focus on the general domain, whereas ours is conducted in the scientific domain.

Overall, our work highlights the limitations of full-document representations for listwise reranking in scientific retrieval. And we leverage IE-based features as document representations for reranking. This makes our contribution novel and significant within the reranking literature.

\begin{figure*}[!ht]
    \centering
    \includegraphics[width=0.30\linewidth]{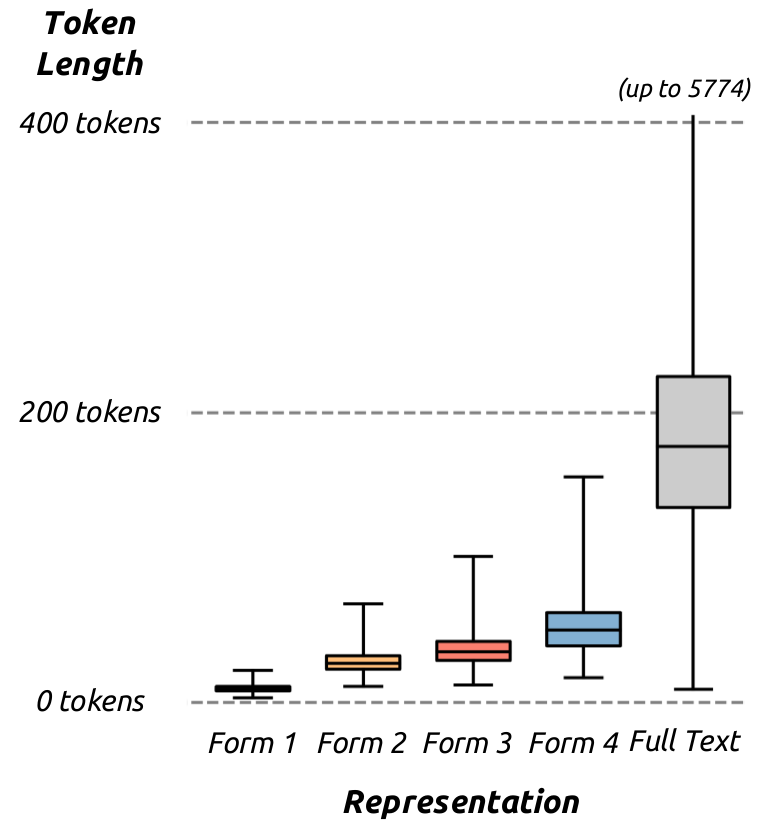}
    \hfill
    \includegraphics[width=0.35\linewidth]{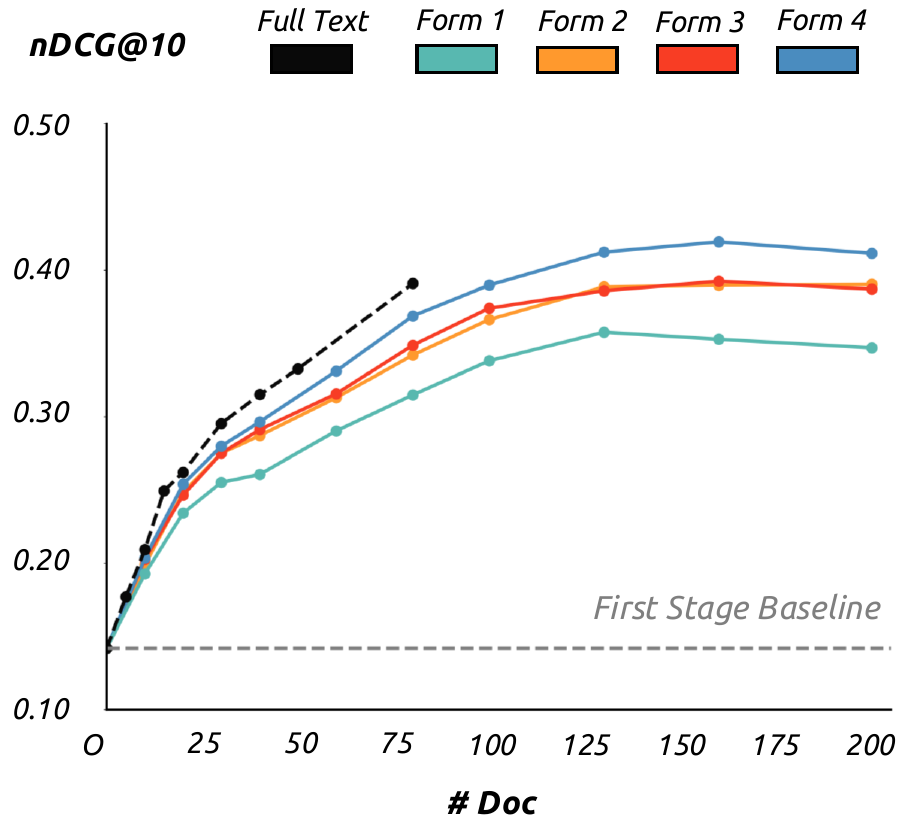}
    \hfill
    \includegraphics[width=0.30\linewidth]{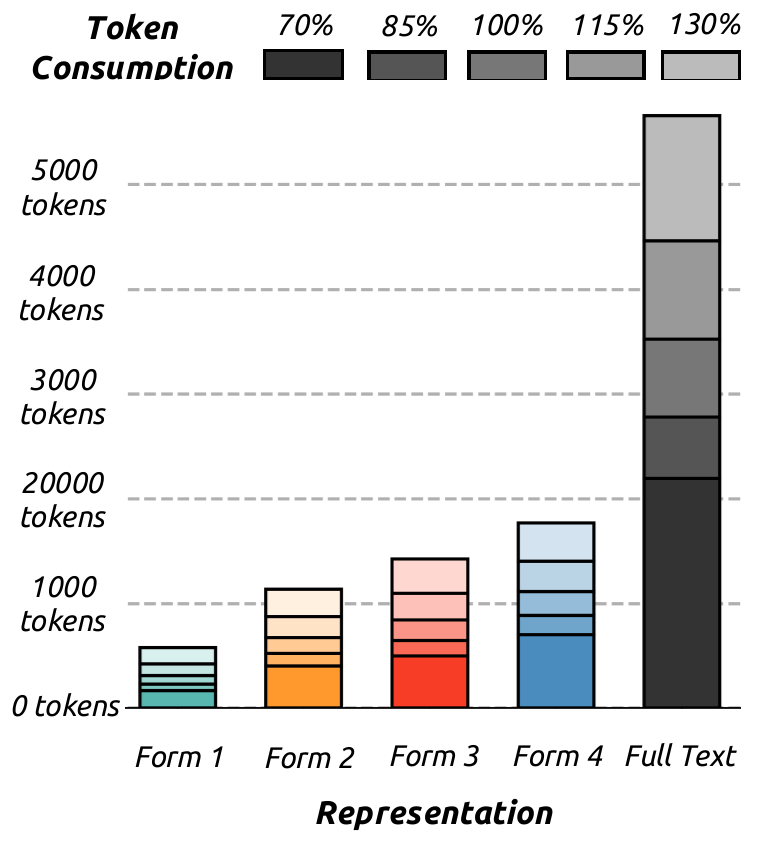}
    \caption{Token efficiency and performance comparison across different document representations. (a) Per-document token length distribution. (b) nDCG@10 scores for different numbers of documents in the context window of single LLM input (c) Context token overhead to reach equal reranking performance.}
    \Description{Composite figure with three plots. 
    (a) Violin/box plots show token counts per document: full text averages around 200 tokens, with a long tail past 5{,}000; four feature-based forms (Form 1–4) are far shorter, roughly 10–50 tokens each. 
    (b) Line chart of nDCG@10 as candidate-set size rises from 10 to 200. Full text reranking leads up to ~80 documents, but exceeds the 32k-token limit thereafter; Form 4 continues improving and overtakes full text beyond 100 candidates. 
    (c) Bar chart of total context tokens required to reach 70\%, 85\%, 100\%, 115\%, and 130\% of the 20-doc full text baseline. Full text needs ≈3{,}500 tokens for 100\% performance, whereas Form 4 reaches the same level with ≈1{,}000 tokens and achieves 130\% performance with fewer tokens than full text at 100\%. 
    The visual message: feature-based representations are dramatically more token-efficient, allowing more documents per prompt and equal or better reranking quality.}

    \label{fig:scaling_results}
    \vspace{-1em}
\end{figure*}

\section{Preliminary Analysis}
\label{section: pre-ex results and analysis}
In this section, we first define the task of listwise reranking, then show the limitations of current full-document-based methods and explore compact and informative document representation options for reranking in the scientific domain.

\subsection{Problem Definition}
\label{sec:problem_definition}

\para{Reranking Input.}
Given a query $q$ and a large corpus $\mathcal{P}_n$ of $n$ scientific papers, a first-stage retriever selects a ranked list of $m$ candidate documents $\mathcal{C} = \{p_{1}, p_{2}, \dots, p_{m}\} \subset \mathcal{P}_n$. Typically we have $m \ll n$ due to the scale of the corpus and the finite capacity of rerankers.
This candidate list $\mathcal{C}$ is then passed to a reranking model, which aims to reorder the documents such that more relevant ones are ranked higher.

\para{Reranking Objective.} 
The goal of document reranking is to find a permutation \(\sigma:\{1,\dots ,m\}\!\to\!\{1,\dots ,m\}\) whose application
\[
\mathcal{C}' \;=\; [\,p_{\sigma(1)},\,p_{\sigma(2)},\dots ,p_{\sigma(m)}\,]
\]
orders documents in more accurate \emph{descending} relevance to \(q\).
The quality is often~\citep{sun2023chatgpt, gangi-reddy-etal-2024-first, liu2024slidingwindowsendexploring} evaluated using top-\(k\) metrics such as nDCG, with small \(k\) values that stress accuracy at the top of the ranked list.

\para{LLM Listwise Reranking.}
Large language models can inspect \(m\) candidate documents in their context window and generate a permutation that reflects their relevance ordering. Formally, the model acts as follows:
\[
\sigma_{\text{LLM}}
   \;=\;
   \mathrm{LLM}(q,\,\mathcal{C})
\]
Applying this permutation to the candidate list yields  
\[
\mathcal{C}'
   \;=\;
   \bigl[
     p_{\sigma_{\text{LLM}}(1)},\;
     p_{\sigma_{\text{LLM}}(2)},\;
     \dots,\;
     p_{\sigma_{\text{LLM}}(m)}
   \bigr].
\]

\subsection{Current Limitations}
\label{sec: limitations}
The standard approach in LLM-based listwise reranking~\citep{sun2023chatgpt, pradeep2023rankvicunazeroshotlistwisedocument, pradeep2023rankzephyreffectiverobustzeroshot, gangi-reddy-etal-2024-first, liu2024slidingwindowsendexploring} feeds the \textit{full text} of each candidate into the model’s context window. While this allows LLMs to capture the complete content of each document, it faces some challenges in the scientific domain.

\para{Suboptimal First-Stage Retrieval.}
Unlike in general-domain scenarios, first-stage retrievers—whether sparse~\citep{robertson2009probabilistic, Nogueira2019FromDT} or dense~\citep{izacard2022unsuperviseddenseinformationretrieval, Wang2022TextEB}—struggle to generalize to the scientific domain~\citep{thakur2021beirheterogenousbenchmarkzeroshot, Bonifacio2022InParsUD}. This is due to the complexity of long-tail concepts~\citep{kang2024taxonomy, kang2025improvingscientificdocumentretrieval} and the lack of large-scale supervised training data~\citep{Bonifacio2022InParsUD, Li2023SAILERSP}. Consequently, their retrieval performance is limited, and truly relevant documents are often ranked much lower~\citep{ajith2024litsearchretrievalbenchmarkscientific}. 

\para{Token Consumption.}
Full text representation also introduces significant token overhead, with individual scientific papers often consuming hundreds or even thousands of tokens~\citep{thakur2021beirheterogenousbenchmarkzeroshot, ajith2024litsearchretrievalbenchmarkscientific, mysore2021csfcubetestcollection}. At the same time, LLMs have a limited effective context length~\citep{Dai2019TransformerXLALA, Xiong2023EffectiveLSA} and are known to suffer from issues such as positional bias for long inputs~\citep{liu2023lostmiddlelanguagemodels, Tian2024DistanceBR}. As a result, full text rerankers are constrained to operate on a small number of candidates, typically 20 documents per input prompt~\citep{sun2023chatgpt, ma2023zeroshotlistwisedocumentreranking, pradeep2023rankvicunazeroshotlistwisedocument, pradeep2023rankzephyreffectiverobustzeroshot}.

When first-stage retrieval is suboptimal and reranking operates over a narrow candidate set, the overall reranking performance is significantly constrained.

\subsection{Semantic Features as Alternatives}
\label{sec: semantic document features}

\subsubsection{Semantic Feature Construction}

Given the limitations of full text representations in scientific-domain reranking, we explore an alternative approach, which is based on document-level semantic features such as categories and keywords. The intuition is that these features are both significantly more concise and capable of preserving the core semantics. Therefore, more candidate documents can be effectively considered during reranking. Specifically, we examine the following four types of document semantic features:

\para{Category.}
Categories~\citep{Sun2023TextCVA, zhang2024teleclass} offer a high-level topical overview of scientific documents. Each document is assigned a three-level hierarchical category path in the format \{\textit{Category}\} → \{\textit{Subcategory}\} → \{\textit{Subsubcategory}\}, which provides a broad-to-specific classification on its topic. 

\para{Section.}
Sections~\citep{Zhou2023LLMADA} consist of multiple subtitle-style strings, each summarizing a major part of the scientific document. They capture the internal structure of the document and serve as mid-level semantic signals that enhance category-level summaries.

\para{Keyword.}
Keywords~\citep{rose2010automatic, Lee2023TowardKGA} are terms or entities that represent fine-grained lexical concepts within a document. They provide the most specific information among different granularities. 

\para{Pseudo Query.}
Pseudo queries~\citep{Sachan2022ImprovingPRA, kang2025improvingscientificdocumentretrieval} are synthetic user questions based on the content. This feature offers a unique query-aligned perspective of the document, simulating how the document might be retrieved in real-world scenarios.


These four features represent the most common types of document-level information extraction (IE)~\citep{Niklaus2018ASOA, wimalasuriya2010ontology}, varying in both granularity and style. All semantic features are extracted using LLM-based \textit{information extraction} methods~\citep{yuan-etal-2023-zero, zhang-etal-2023-aligning}. We adopt zero-shot prompting with instruction-tuned models such as GPT-4.1~\citep{openai2024gpt41} and the Qwen model family~\citep{qwen2025qwen25technicalreport}. This approach leverages the inherent knowledge of pretrained language models~\citep{min2023recent, li2024pre} to extract information without requiring task-specific training~\citep{yuan-etal-2023-zero}, while demonstrating strong generalization across domains and schema variations~\citep{xu2025zeroshotopenschemaentitystructure}.

Among these features, \textit{sections}, \textit{keywords}, and \textit{pseudo queries} have multiple elements for each document. To reduce noise and focus on the most relevant content, we apply an adaptive selection strategy during inference time. We compute dense embedding similarities between the query and each extracted element and retain only the most relevant: 5 keywords (from 30), 1 pseudo query (from 20), and 1 section (from 3). This improves content relevance while minimizing token overhead.

However, relying on a single feature often lacks representational power. For example, using only the category tends to be too coarse-grained and offers limited discriminative ability, while keywords alone may lack sufficient context or background information. To address this, we represent each document using \textit{combinations} of features. Specifically, we explore the following four configurations of different overall granularities. 

\begin{enumerate}[label=\textbf{Form~\arabic*.}, align=left, leftmargin=30pt, itemsep=4pt, parsep=0pt, topsep=4pt, partopsep=0pt]
\item \textit{Pseudo Query}
\item \textit{Category}
\item \textit{Category + Section}
\item \textit{Category + Section + Keywords}
\end{enumerate}

Ablation studies on the effect of each specific component can further be found in Section~\ref{sec: ablation}. The overview of representation construction is presented in Figure~\ref{fig:representation_pipe}.

\subsubsection{Empirical Validation}

\label{sec: prelim exp setup}

To evaluate the effectiveness of our feature-based representations, we conduct a series of experiments on the LitSearch~\citep{ajith2024litsearchretrievalbenchmarkscientific}, using GPT-4.1-mini~\citep{openai2024gpt41} as the reranking backbone. Semantic features are extracted using Qwen3-8B~\citep{qwen2025qwen25technicalreport}. Following~\citet{gangi-reddy-etal-2024-first}, we use Contriever~\citep{izacard2022unsuperviseddenseinformationretrieval} as the first-stage retriever and the similarity encoder for adaptive selection.

\para{Token Length Per Document.}
We first evaluate the token cost of different representations. As shown in Figure~\ref{fig:scaling_results} (a), full text inputs consume around 200 tokens per document on average, with some exceeding 5,000 tokens, limiting the number of documents that can fit into the context window. In contrast, semantic-feature-based representations (Forms 1-4) are much more compact, each averaging between 10 and 50 tokens.

\para{Number of Documents in a Single Input.}
We then evaluate how different representations perform as the number of candidate documents increases (up to 200) within a 32k-token context window. As shown in Figure~\ref{fig:scaling_results} (b), all representations improve as more documents are included. In the range of 0 to 80 candidates, full text representations perform better than feature-based ones. This is expected, since full text retains complete document information, though at a much higher token cost. However, beyond 80 candidates, full text inputs frequently exceed the context length limit. In contrast, feature-based methods, especially Form 4, remain compact and can handle more candidates without issue, while still achieving comparable performance.

\para{Token Cost for Equal Reranking Performance.}
Finally, we examine how many context tokens are needed to achieve the same reranking performance across different representations. Using results from the previous two experiments, we estimate the total token cost by combining the number of documents required with the average token count per document. We take full text reranking with 20 documents as the 100\% performance baseline, and compare how many tokens each method needs to reach 70\%, 85\%, 100\%, 115\%, and 130\% of that level. As shown in Figure~\ref{fig:scaling_results} (c), compact feature-based representations can match full text performance with significantly fewer tokens. For instance, reaching 100\% performance requires around 3,500 tokens for full text, but only 1,000 or even a few hundred tokens for feature-based inputs.

\subsubsection{Discussion}

The results above demonstrate that feature-based document representations offer clear advantages over full text inputs in reranking. They can express the core content of a document using far fewer tokens, allowing significantly more retrieved candidates to be included within the same token overhead. This is especially valuable for recovering relevant documents that were assigned lower scores by the first-stage retriever. Considering both token efficiency and reranking performance, we finally select Form 4 as our feature-based representation.

However, in the second set of experiments where the number of documents is held constant, we also observe that feature-based representations may underperform full text. After all, offline information extraction is \textit{not} perfect. The process may omit subtle but important cues present in the full text.

Therefore, these results motivate a balanced reranking strategy: using semantic features in the early stage to cover a broader range of candidates, followed by a refinement stage that reranks the top candidates using full text inputs for more in-depth comparison.

\section{Methodology}

Based on the preliminary analysis, we propose a new three-stage reranking framework in the scientific domain: (i) offline extraction of structured semantic features; (ii) coarse-grained reranking using the features to select a subset of top candidates; and (iii) fine-grained reranking over the subset with full text inputs.

\subsection{Compact and Informative Representation}

To construct compact, feature-based representations for each scientific document, we first perform document-level information extraction offline. As described in Section~\ref{sec: semantic document features}, we use LLMs to obtain target semantic features: \textit{category}, \textit{section}, and \textit{keyword}. Formally, for a given document $p_i$, we obtain:

$$
\text{LLM}(p_i)
=
\text{Category}_i,\, [\text{Section}_i^{j}],\, [\text{Keyword}_i^{j}]
$$

These features are designed to capture the core semantics of scientific documents in a token-efficient manner. The extraction is performed offline, and the results are cached and reused during inference time to avoid any additional runtime delay. 

\subsection{Coarse-grained Reranking w. Compact Features}

Given a query \(q\), we obtain a document relevance ranking \(\mathcal{C} = [p_1,\, p_2,\, \dots,\, p_m]\) from a first-stage retriever. We then replace each document with a compact semantic representation and apply LLM-based listwise reranking to identify a high-quality subset.

For sections and keywords, we apply \textit{adaptive selection} to select only those most relevant to the query. Specifically, for each unit (single section or keyword) \(u\), we compute the cosine similarity with text embedding and select the top-5 keywords and the most relevant section. (Analysis on the number of keyword used can be found in Section~\ref{sec: keyword num}.) The final feature-based representation \(r_i\) is formed as:
\[
r_i = \text{Concat}\left[\text{Category}_i,\, \text{Section}_i^*,\, \text{Keywords}_i^*\right]
\]
where \(^*\) indicates adaptively selected elements.

With the feature-based representations \(\mathcal{R} = [r_1,\, r_2,\, \dots,\, r_m]\), we then use the LLM to perform listwise reranking, producing a permutation \(\sigma_{\text{feat}} = \text{LLM}(q,\, \mathcal{R}) \in S_m\). Applying this permutation to the original candidate list \(\mathcal{C} = [p_1,\, p_2,\, \dots,\, p_m]\) yields the reranked output for the coarse-grained reranking:
\[
\mathcal{C}_{\text{feat}} = [p_{\sigma_{\text{feat}}(1)},\, p_{\sigma_{\text{feat}}(2)},\, \dots,\, p_{\sigma_{\text{feat}}(m)}]
\]
We then keep the top-\(k\) documents from this list to form the seed set for full text reranking:
\[
\mathcal{C}_{\text{seed}} = [p_{\sigma_{\text{feat}}(1)},\, \dots,\, p_{\sigma_{\text{feat}}(k)}], \quad k < m
\]

Coarse-grained reranking with compact document representations greatly expands the number of candidates that can be considered in the same LLM input. This broader coverage helps recover scientific documents that were assigned low scores by the first-stage retriever.

\subsection{Fine-grained Reranking w. Full Text}

In the second reranking stage, we refine the ranking over the seed set of candidates using full text representation. Specifically, we take the seed set \(\mathcal{C}_{\text{seed}} = [p_1',\, p_2',\, \dots,\, p_k']\), (we use $'$ to distinguish them from the first-stage input) obtained from the previous stage, and replace each compact representation with its original document.

Let \(\mathcal{T} = [t_1,\, t_2,\, \dots,\, t_k]\) denote the full text of the selected documents, where \(t_i\) corresponds to the full text content of passage \(p_i'\). We then once again use LLM to perform listwise reranking:
\[
\sigma_{\text{text}} = \text{LLM}(q,\, \mathcal{T}) \in S_k
\]
With the permutation \(\sigma_{\text{text}}\) we get the final ranking:
\[
\mathcal{C}_{\text{final}} = [p_{\sigma_{\text{text}}(1)}',\, \dots,\, p_{\sigma_{\text{text}}(k)}']
\]

Fine-grained reranking with full text inputs recovers details that may be lost during the information extraction process. Since the candidate set has already been narrowed down, full text representation can now be used without exceeding the LLM’s context limit.

Overall, our strategy effectively addresses the challenges of LLM-based listwise reranking in the scientific domain. The coarse reranking stage expands the reranking pool, improving robustness to suboptimal first-stage retrieval in the scientific domain. The fine-grained reranking stage, in turn, preserves sufficient detail for precise relevance comparisons.
% ...

\begin{table*}[!ht]
\caption{Reranking performance of different reranking methods on LitSearch~\citep{ajith2024litsearchretrievalbenchmarkscientific}, CSFCube~\citep{mysore2021csfcubetestcollection}, NFCorpus~\citep{boteva2016full}, SciFact~\citep{wadden-etal-2020-fact} and Trec-Covid~\citep{trec-covid} with nDCG@10 (N@10) and Recall@10 (R@10) as metrics. Higher scores under the same setting are bold. All values are shown as percentages.}
\centering
\begin{tabular}{ll*{10}{c}}
\toprule
\multirow{2}{*}{\textbf{Reranking Model}} &
\multirow{2}{*}{\textbf{Reranking Strategy}} &
\multicolumn{2}{c}{\textbf{LitSearch}} &
\multicolumn{2}{c}{\textbf{CSFCube}} &
\multicolumn{2}{c}{\textbf{NFCorpus}} &
\multicolumn{2}{c}{\textbf{SciFact}} &
\multicolumn{2}{c}{\textbf{Trec-Covid\footnotemark}} \\
\cmidrule(lr){3-4} \cmidrule(lr){5-6} \cmidrule(lr){7-8} \cmidrule(lr){9-10} \cmidrule(lr){11-12}
& & N@10 & R@10 & N@10 & R@10 & N@10 & R@10 & N@10 & R@10 & N@10 & R@10 \\
\midrule
None & First-stage retrieval only & 14.2 & 21.2 & 23.6 & 16.4 & 32.8 & 15.5 & 67.7 & 81.3 & 59.6 & 1.6 \\
\midrule
RankMistral & Vanilla Reranking$\ _{Full}$    & 20.1 & 24.3 & 21.9 & 13.3 & 22.9 & 9.5 & 16.9 & 27.3 & 52.3 & 1.3 \\
RankVicuna  & Vanilla Reranking$\ _{Sliding}$ & 26.5 & 32.4 & 28.7 & 17.5 & 26.3 & 11.6 & 29.4 & 40.7 & 64.1 & 1.7 \\
RankZephyr  & Vanilla Reranking$\ _{Sliding}$ & 34.1 & 39.0 & 34.1 & 22.7 & 28.6 & 15.1 & 44.4 & 80.0 & 59.5 & 1.6 \\
\midrule
\multirow{2}{*}{Qwen3-32B}
& Vanilla Reranking$\ _{Full}$   & 25.7 & 27.3 & 28.4 & 20.2 & 36.8 & 17.2 & 77.6 & 85.7 & 67.0 & 1.7 \\
& \Method$\ _{Full}$    & \textbf{39.6} & \textbf{41.9} & \textbf{31.8} & \textbf{23.4} & \textbf{39.8} & \textbf{18.5} & \textbf{80.8} & \textbf{89.1} & \textbf{76.4} & \textbf{2.0} \\
\midrule
\multirow{2}{*}{Qwen3-32B}
& Vanilla Reranking$\ _{Sliding}$ & 39.8 & 43.5 & 31.2 & 22.9 & 38.8 & 18.0 & 80.0 & 89.8 & 76.7 & 2.1 \\
& \Method$\ _{Sliding}$  & \textbf{42.2} & \textbf{46.2} & \textbf{35.5} & \textbf{26.6} & \textbf{39.8} & \textbf{18.7} & \textbf{81.7} &\textbf{ 90.3} & \textbf{78.1} & 2.1 \\
\midrule
\multirow{2}{*}{Gemini 2.0 Flash}
& Vanilla Reranking$\ _{Full}$   & 26.1 & 27.3 & 28.5 & 18.8 & 35.9 & 16.8 & 75.6 & 85.4 & 68.9 & 1.8 \\
& \Method$\ _{Full}$    & \textbf{40.7} & \textbf{44.2} & \textbf{35.1} & \textbf{27.8} & \textbf{38.0} & \textbf{17.5} & \textbf{77.6} & \textbf{88.3} & \textbf{76.1} & \textbf{2.0} \\
\midrule
\multirow{2}{*}{Gemini 2.0 Flash}
& Vanilla Reranking$\ _{Sliding}$ & 40.8 & 44.6 & 32.2 & 23.2 & 38.3 & 17.9 & 77.2 & 88.8 & \textbf{79.7} & 2.2 \\
& \Method$\ _{Sliding}$  & \textbf{43.3} & \textbf{48.9} & \textbf{38.1} & \textbf{31.0} & \textbf{39.0} & \textbf{18.6} & \textbf{78.0} & \textbf{91.1} & 79.2 & 2.2 \\
\midrule
\multirow{2}{*}{GPT-4.1-mini}
& Vanilla Reranking$\ _{Full}$   & 26.1 & 27.2 & 28.7 & 19.7 & 37.1 & 16.8 & 76.7 & 84.9 & 68.4 & 1.8 \\
& \Method$\ _{Full}$    & \textbf{46.3} & \textbf{49.4} & \textbf{39.0} & \textbf{30.5} & \textbf{41.4} & \textbf{19.1} & \textbf{79.6} & \textbf{89.8} & \textbf{80.7} & \textbf{2.2} \\
\midrule
\multirow{2}{*}{GPT-4.1-mini}
& Vanilla Reranking$\ _{Sliding}$ & 41.9 & 44.4 & 34.9 & 25.3 & 40.0 & 17.9 & 79.3 & 90.5 & 81.0 & 2.2 \\
& \Method$\ _{Sliding}$  & \textbf{45.8} & \textbf{49.1} & \textbf{39.4} & \textbf{30.4} & \textbf{41.4} & \textbf{19.2} & \textbf{80.4} & \textbf{91.6} & \textbf{80.5} & 2.2 \\
\bottomrule
\end{tabular}

\label{tab:evaluated_models}
\vspace{-1.0em}
\end{table*}

\section{Experiments}
\label{sec: experiments}

In this section, we first outline the experimental setup, then present the main reranking performance results, followed by ablation studies and further analysis.

\subsection{Experimental Setup}
\label{sec: experimental setup}

\para{Datasets \& Metric.}
We evaluate different reranking methods on five high-quality scientific retrieval benchmarks: LitSearch~\citep{ajith2024litsearchretrievalbenchmarkscientific}, CSFCube~\citep{mysore2021csfcubetestcollection}, NFCorpus~\citep{boteva2016full}, SciFact~\citep{wadden-etal-2020-fact} and Trec-Covid~\citep{trec-covid}, which collectively cover diverse scientific domains including computer science, biomedical and clinical information retrieval. Follow standard practice~\citep{sun2023chatgpt, ma2023zeroshotlistwisedocumentreranking, gangi-reddy-etal-2024-first, liu2024slidingwindowsendexploring}, we report nDCG and Recall as our evaluation metrics for top-10 document. 

\para{Models.}
For the reranking backbones, we use Qwen3-32B~\citep{qwen2025qwen25technicalreport} as the representative open-source model, Gemini 2.0 Flash~\citep{google2025gemini} and GPT-4.1-mini~\citep{openai2024gpt41} as proprietary examples. For generative parameters, we use a temperature of 1.0 and a fixed random seed of 42 for reproducibility. 

For semantic feature extraction, we find that small open-source models are sufficiently effective; we therefore adopt Qwen3-8B~\citep{qwen2025qwen25technicalreport} for document feature extraction. To further assess the quality of information extraction outcome, we also conducted a human evaluation process shown in Section~\ref{human evaluation}. For both first-stage retrieval and semantic similarity scoring for features filtering, we use Contriever~\citep{izacard2022unsuperviseddenseinformationretrieval} following previous works~\citep{reddy2024refitrelevancefeedbackreranker, gangi-reddy-etal-2024-first}. To ensure the robustness against different first-stage retrievers, We also evaluate the performance of our method with diverse retrievers in Section~\ref{first-stage retrievers}.

\para{Baselines.}
Our method is a \textit{zero-shot}, \textit{model-agnostic} approach. Therefore, we primarily compare against prior reranking baselines under the same setting, most notably vanilla listwise reranking~\citep{sun2023chatgpt} with zero-shot instruction-following. We also consider the sliding window strategy~\citep{sun2023chatgpt, ma2023zeroshotlistwisedocumentreranking}, a test-time scaling technique~\citep{xia2025generativeaiactii} designed to expand the reranking context. Results are reported both with and without the sliding window strategy. We conduct a focused comparison between our method and the sliding window strategy in Section~\ref{sec: further analysis}.

We also include some reranking methods from the literature that represent the-state-of-the-art supervised LLM rerankers like RankVicuna~\citep{pradeep2023rankvicunazeroshotlistwisedocument}, RankZephyr~\citep{pradeep2023rankzephyreffectiverobustzeroshot}, and RankMistral~\citep{liu2024slidingwindowsendexploring}, which are trained on large-scale general-domain datasets such as MS MARCO~\citep{bajaj2018msmarcohumangenerated}. While these models differ from ours in both model size and supervision conditions, we report them for reference to illustrate the overall performance distribution.

For a fairer comparison with supervised baselines, we evaluate our method using smaller-parameter LLMs (7B–8B) in Section~\ref{reranking backbone}, allowing us to isolate the effect of model size from supervision.

\para{Reranking Parameters.}
We follow the standardized setup~\citep{sun2023chatgpt, ma2023zeroshotlistwisedocumentreranking, pradeep2023rankvicunazeroshotlistwisedocument, pradeep2023rankzephyreffectiverobustzeroshot, gangi-reddy-etal-2024-first} from prior work to unify reranking parameters. For vanilla reranking, we include 20 full text documents within the context. An exception is RankMistral~\citep{liu2024slidingwindowsendexploring}, which benefits from long-context training and is able to process 100 full text documents in a single input. When using the sliding window strategy, we rank a total of 100 full text documents by applying a window size of 20 with a step size of 10. For \Method, we include 200 compact representations in the coarse reranking stage, followed by 20 full text documents in the fine-grained reranking stage.

\subsection{Main Results}
\footnotetext{Low because each query in Trec-Covid has far more than 10 relevant documents.}

\subsubsection{Reranking Performance.}
We begin by evaluating the performance of different reranking methods on the target academic retrieval benchmarks, as shown in Table~\ref{tab:evaluated_models}.

First, we observe that all reranking methods outperform the first-stage dense retriever baseline, confirming the critical role of reranking in scientific document retrieval. However, while supervised models such as RankVicuna~\citep{pradeep2023rankvicunazeroshotlistwisedocument} do yield noticeable improvements, their gains remain limited, partly because these models are relatively small (7B–8B parameters) and thus less competitive despite having supervision. Meanwhile, current general LLMs already demonstrate strong instruction following capabilities for zero-shot reranking. Furthermore, our proposed method, \Method, which is model-agnostic, and training-free, delivers even stronger and consistent gains across different LLM backbones. Without using sliding windows, \Method\ improves the average nDCG@10 from 47.2 to 54.9, and when combined with sliding window inputs, it still provides an average gain of +2.0 nDCG@10 (from 54.1 to 56.2).

We also find that both \Method\ and the sliding window strategy independently lead to substantial performance gains (ours being larger and more efficient; see Section~\ref{sec: further analysis}). This is likely because both methods expand the reranking scope, which is particularly valuable in the scientific domains considering the suboptimal first-stage retrieval. These findings also reinforce our earlier analysis of current limitations presented in Section~\ref{sec: limitations}.

\subsubsection{Ablation Studies.}
\label{sec: ablation}
To evaluate the contribution of each component in our design, we conduct ablation studies on the LitSearch~\citep{ajith2024litsearchretrievalbenchmarkscientific} dataset using GPT-4.1-mini~\citep{openai2024gpt41} as the backbone.

Specifically, we remove individual semantic features—\textit{category}, \textit{section}, and \textit{keywords}—from the compact representation to assess their relative importance. In addition, we ablate two key components of our pipeline: \textit{adaptive selection} and \textit{fine-grained reranking}, to understand their impact on overall performance.

\begin{figure}[!ht]
    \centering
    \includegraphics[width=0.97\linewidth]{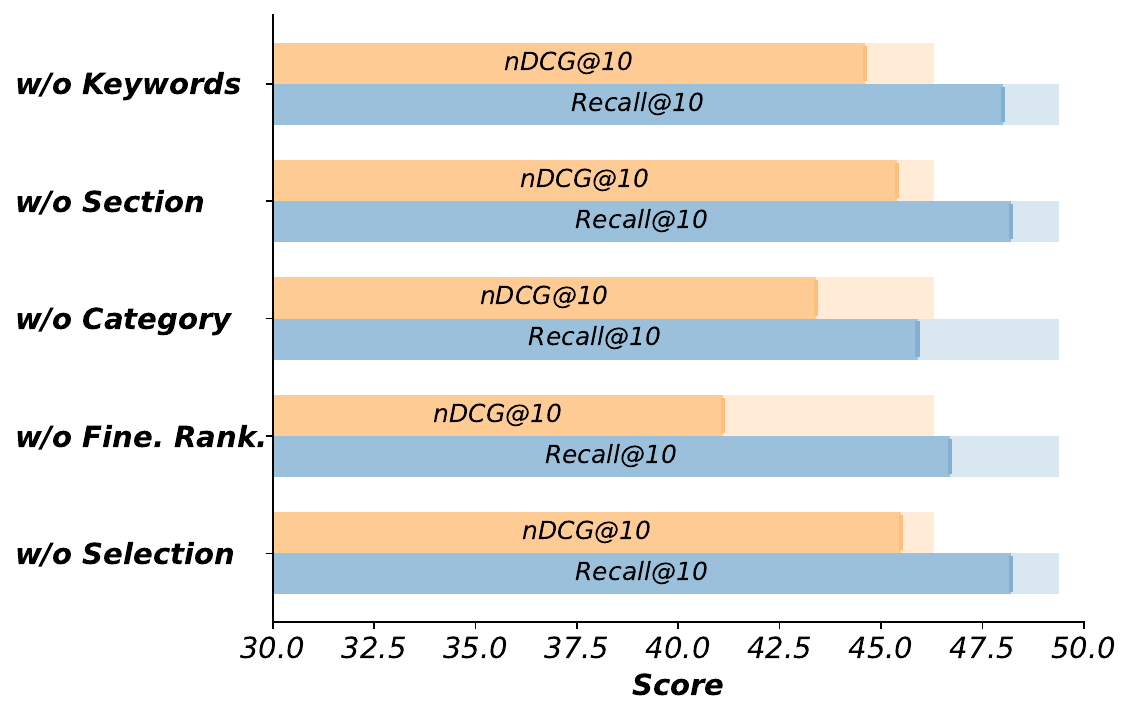}
    \caption{Ablation study results. Foreground bars report performance after removing the labeled component; the translucent background bars indicate the full method score. }
    \Description{A horizontal bar chart shows ablation results on the LitSearch dataset (backbone: GPT-4.1-mini). The x-axis is “Score” from roughly 30 to 50. There are five rows labeled: w/o Keywords, w/o Section, w/o Category, w/o Fine. Rank. (removing fine-grained reranking), and w/o Selection (removing adaptive selection). For each row, two solid foreground bars report the metrics nDCG@10 (blue) and Recall@10 (red). Behind each, a faint background bar of the same color marks the full model (no ablation) score for reference. In every case the solid bars are shorter than the background bars, meaning performance drops when the component is removed; the largest drops are for removing Selection and Fine. Rank., while removing Category, Section, or Keywords yields smaller but consistent declines.}
    \label{fig:ablation features}
\end{figure}

As shown in Figure~\ref{fig:ablation features}, removing any single semantic component (category, section, or keywords) consistently degrades reranking performance, confirming that each feature contributes complementary information. Category captures coarse-grained background context that tends to align more easily with the query but offers less specific detail. In contrast, section and keywords reflect finer-grained semantics that are harder to match but, when relevant, provide highly targeted signals. This balance shows the value of combining semantic features across different levels of abstraction.

Additionally, we find that both adaptive selection and fine-grained reranking contribute to the final performance. Adaptive selection ensures that the most query-relevant features are retained, while the fine-grained reranking stage substantially compensates for potential information loss in compact representations by recovering more nuanced evidence from the full text. 

\subsection{Further Analysis}

\subsubsection{Hyperparameter Study}

We investigate the effect of key hyperparameters in \Method\ using the GPT-4.1-mini~\citep{openai2024gpt41} reranker on the LitSearch~\citep{ajith2024litsearchretrievalbenchmarkscientific} dataset, reporting nDCG@10, Recall@10, token usage, and estimated API cost\footnote{Based on the pricing listed on the official \href{https://platform.openai.com/docs/pricing}{OpenAI API Platform website}.}. Our analysis focuses on understanding the \textit{trade-off} between cost and performance, as well as the \textit{robustness} of \Method\ across a wide hyperparameter range.

\para{Number of Keywords.} By default, \Method\ extracts 30 keywords from each document and selects the top 5 most relevant ones via cosine similarity in the embedding space, concatenating them into the document representation for \textit{ coarse-grained} reranking. We vary the number of concatenated keywords from 0 to 20 and report the immediate results from coarse-grained reranking (Table~\ref{tab:num of kw}).

\begin{table}[!ht]
\caption{Effect of the number of selected keywords on coarse-grained reranking: performance, token usage and cost.}
\centering
\begin{tabular}{lcccc}
\toprule[1pt]
\textbf{\# Keyword} & 
\textbf{N@10} & 
\textbf{R@10} & 
\makecell{\textbf{Token}\\\textbf{Usage}} & 
\textbf{Cost} \\
\midrule[0.7pt]
0 keyword& 38.9 & 44.2 & 1.80M & \$0.72 \\
1 keyword& 40.4 & 46.3 & 1.95M & \$0.78 \\
3 keywords& 40.1 & 44.9 & 2.26M & \$0.90 \\
5 keywords & 41.1 & 46.7 & 2.57M & \$1.03 \\
10 keywords & 42.4 & 46.5 & 3.33M & \$1.33 \\
15 keywords & 42.6 & 48.6 & 4.10M & \$1.64 \\
20 keywords & 42.4 & 47.4 & 4.87M & \$1.95 \\
\bottomrule[1pt]
\end{tabular}

\label{tab:num of kw}
\end{table}

Results show a clear \textit{cost--performance trade-off}: adding more keywords linearly increases token usage while steadily improving coarse-grained reranking quality. However, performance gains diminish beyond 10 keywords, as later keywords contribute weaker relevance signals. Importantly, \Method\ achieves strong performance even with substantially fewer keywords, indicating that it is \textit{robust} to this hyperparameter and does not require precise tuning to remain competitive.

\para{Fine-Grained Pool Size.}
\label{sec: keyword num}
In the main experiments, the fine-grained reranker processes the top 20 candidates from the coarse stage to align with other baselines. Here, we vary the candidate pool from 5 to 100 documents and report the reranking performance of the final fine-grained stage (Table~\ref{tab:topk_fine_rerank}).

\begin{table}[!ht]
\caption{Effect of the fine-grained candidate pool size: performance, token usage and cost.}

\centering

\begin{tabular}{lcccc}
\toprule[1pt]
\textbf{Pool Size} & 
\textbf{N@10} & 
\textbf{R@10} & 
\makecell{\textbf{Token}\\\textbf{Usage}} & 
\textbf{Cost} \\
\midrule[0.7pt]
5 docs & 43.0 & 46.7 & 0.25M & \$0.10 \\
10 docs & 44.3 & 46.7 & 0.50M & \$0.20 \\
20 docs & 46.3 & 49.4 & 1.01M & \$0.40 \\
40 docs & 45.4 & 48.0 & 2.01M & \$0.81 \\
60 docs & 45.0 & 48.3 & 3.02M & \$1.21 \\
80 docs & 45.3 & 48.0 & 4.03M & \$1.61 \\
100 docs & 46.0 & 49.1 & 5.03M & \$2.01 \\
\bottomrule[1pt]
\end{tabular}

\label{tab:topk_fine_rerank}

\end{table}

We again observe a \textit{robust cost--performance trade-off}: increasing the pool size up to 20 yields the most significant improvements, after which performance plateaus while token usage continues to grow. This plateau suggests that \Method\ maintains \textit{stable effectiveness} across a broad range of pool sizes, allowing practitioners to adjust this parameter flexibly according to budget without risking notable performance loss.

\para{Summary.}  
These studies confirm that \Method\ offers tunable hyperparameters that enable smooth navigation of cost--performance trade-offs, while maintaining consistently strong results. Its effectiveness is \textit{not narrowly dependent} on exact hyperparameter values, alleviating concerns of fragility or heavy manual tuning. For example, varying the number of keywords between 5 and 20 results in only a 3.1\% variance in nDCG@10, underscoring the robustness of the method.

\subsubsection{Comparison with Sliding Window}
\label{sec: further analysis}

Sliding window strategy~\citep{sun2023chatgpt} is a widely used~\citep{ma2023zeroshotlistwisedocumentreranking, pradeep2023rankvicunazeroshotlistwisedocument, pradeep2023rankzephyreffectiverobustzeroshot, gangi-reddy-etal-2024-first} method for overcoming the context length limitations of LLMs in listwise reranking. Instead of ranking all candidates in a single prompt, it partitions the candidate list into overlapping windows and reranks each window independently from the bottom up. This method has been shown to notably improve reranking performance~\citep{sun2023chatgpt}.

As noted in Section~\ref{sec: experimental setup}, we evaluate both the standalone and combined use of sliding windows with our method. While \Method\ and sliding window are orthogonal and compatible, isolating them in comparison allows us to clearly demonstrate the superior efficiency and effectiveness of our approach when used alone.

\begin{table}[!ht]
\caption{Comparison of performance, token usage, and API cost between \Method \ and Sliding Window.}
\centering
\begin{tabular}{lccccc}
\toprule[1pt]
\textbf{Method} & 
\textbf{N@10} & 
\textbf{R@10} & 
\makecell{\textbf{Token} \textbf{Usage}} & 
\textbf{Cost} \\
\midrule[0.7pt]
Vanilla & 26.0 & 27.3 & 1.01M & \$0.40 \\
Sliding & 40.8 & 44.2 & 9.06M & \$3.62 \\
\Method & 42.2 & 45.2 & 3.60M & \$1.44 \\
Both & 43.8 & 48.1 & 11.65M & \$4.66 \\
\bottomrule[1pt]
\end{tabular}

\label{tab:sliding_vs_feature}
\end{table}

We evaluate \Method\ and the sliding window strategy on LitSearch and CSFCube on three tested models, reporting nDCG@10, total token consumption, and API cost for GPT-4.1-mini. As shown in Table~\ref{tab:sliding_vs_feature}, \Method\ not only achieves better average reranking quality (nDCG@10 improved by 1.4), but also requires only 40\% of the token budget compared to sliding windows. These results demonstrate that \Method\ provides an efficient yet effective alternative for expanding reranking range. Moreover, combining these two methods can further boost performance, offering an advanced solution for complex retrieval scenarios.

\begin{table*}[!ht]
\centering
\caption{Human evaluation results for zero-shot LLM-based information extraction. The outcome achieves high accuracy, faithfulness, and completeness, with competitive preference rates compared to human annotations.}

\begin{tabular}{l >{\raggedright\arraybackslash}m{11cm} l} 
\toprule[1pt]
\textbf{Metric} & \textbf{Definition} & \textbf{Score} \\
\midrule[0.7pt]
Category Accuracy & Proportion of LLM-assigned categories matching human-annotated ground truth & 28/29 (96.6\%). \\
\midrule[0.7pt]
Topic Accuracy & Proportion of LLM-assigned topics matching human-annotated ground truth & 28/29 (96.6\%). \\
\midrule[0.7pt]
Section Faithfulness & 1 - Proportion of generated sections that do not actually appear in the document. & 192/197 (97.5\%) \\
\midrule[0.7pt]
Section Completeness & 1 - Proportion of true sections in the document that the LLM failed to identify. & 177/197 (89.8\%) \\
\midrule[0.7pt]
Keyword Faithfulness & 1 - Proportion of generated keywords that do not actually appear in the document. & 585/623 (93.9\%) \\
\midrule[0.7pt]
Keyword Completeness & 1 - Proportion of true keywords in the document that the LLM failed to identify. & 617/623 (99.0\%) \\
\midrule[0.7pt]
Win Rate & Fraction of times annotators preferred the LLM's annotation over the human's. & 7/15 (46.7\%) \\
\bottomrule[1pt]
\end{tabular}

\label{tab:human-eval}
\end{table*}

\subsubsection{Impact of First-Stage Retrievers}
\label{first-stage retrievers}

\begin{figure}[!ht]
    \centering
    \includegraphics[width=0.97\linewidth]{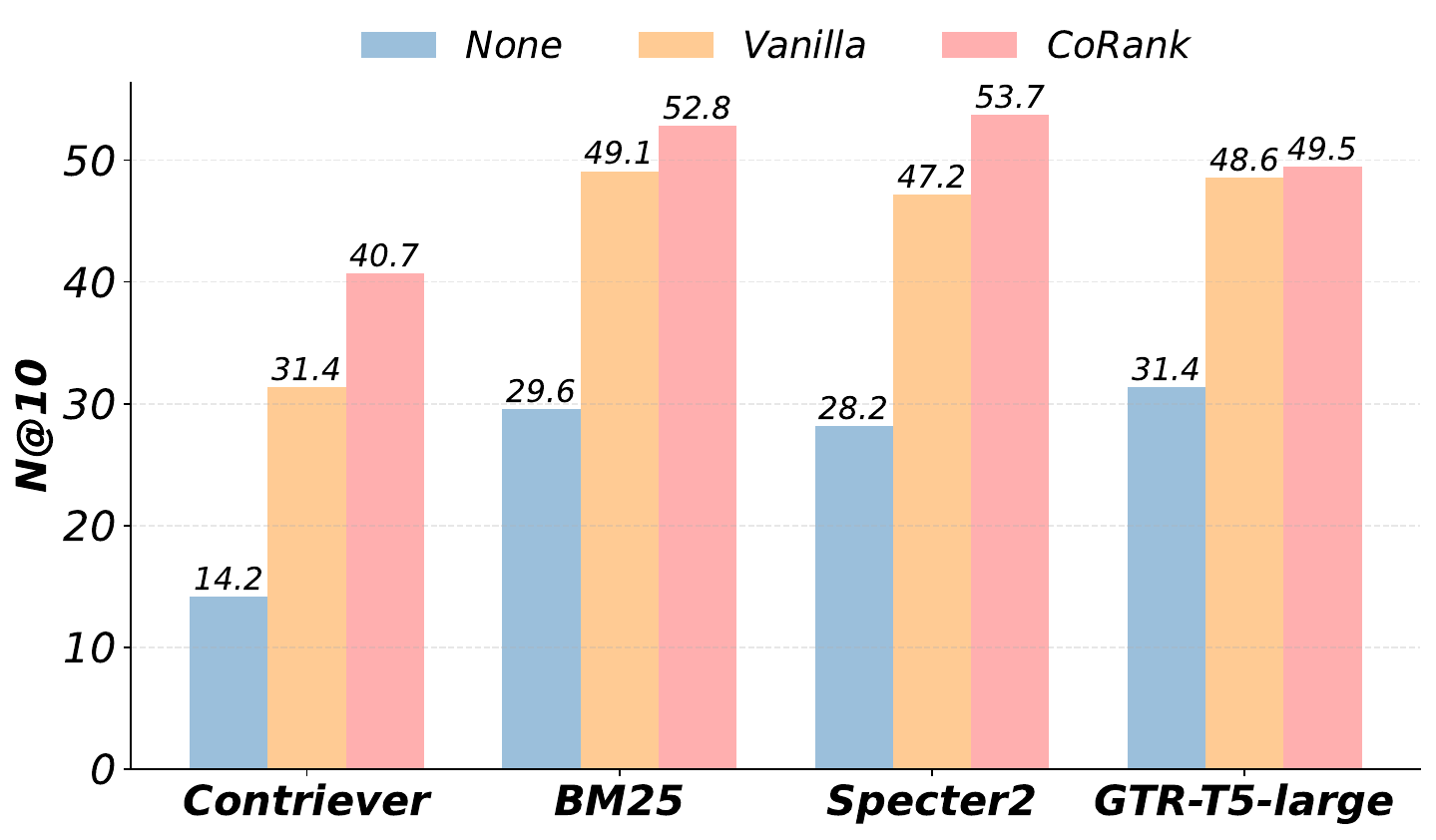}
    \caption{Comparison of reranking methods (None, Vanilla, \Method) across four first-stage retrievers on LitSearch with Gemini 2.0 Flash. }
    \Description{The figure reports N@10 on LitSearch (Gemini 2.0 Flash) for four first-stage retrievers—Contriever, BM25, Specter2, and GTR-T5-large—under three reranking settings: None, Vanilla, and CoRank. For Contriever the scores are 14.2 (None), 31.4 (Vanilla), and 40.7 (CoRank); for BM25 they are 29.6, 49.1, and 52.8; for Specter2 they are 28.2, 47.2, and 53.7; and for GTR-T5-large they are 31.4, 48.6, and 49.5. CoRank is the best method for every retriever, improving over the raw first-stage (“None”) by +26.5, +23.2, +25.5, and +18.1 points respectively, and over the Vanilla reranker by +9.3, +3.7, +6.5, and +0.9. Under CoRank the ordering is Specter2 (53.7) > BM25 (52.8) > GTR-T5-large (49.5) > Contriever (40.7), indicating consistent gains across retrievers and robustness to the first-stage choice.}
    \label{fig:first-stage retriever}
\end{figure}

Following previous works~\citep{reddy2024refitrelevancefeedbackreranker, gangi-reddy-etal-2024-first}, we choose Contriever~\citep{izacard2022unsuperviseddenseinformationretrieval} as our first-stage retriever in our main experiments. To further evaluate robustness to the choice of the first-stage retriever. We here additionally evaluate the performance of our method with BM25~\citep{robertson2009probabilistic}, Specter2~\citep{Singh2022SciRepEvalAM}, GTR-T5-large~\citep{ni-etal-2022-large} as first-stage retrievers. We run evaluation on LitSearch~\citep{ajith2024litsearchretrievalbenchmarkscientific} with Gemini 2.0 Flash~\citep{google2025gemini}. The results are shown in Figure~\ref{fig:first-stage retriever}.

The first-stage retrievers exhibited differences in their standalone retrieval quality. However, our method consistently improved retrieval quality over the vanilla reranking baseline, regardless of which initial retriever was used. In other words, our gains are not confined to weak initial retrieval: \Method \ adds value on strong first-stage retrievers as well, indicating that our method is robust to the choice of first-stage retriever. 

\subsubsection{Impact of Reranking Backbones}
\label{reranking backbone}

To further validate the robustness of our reranking method to different reranking backbones, we conducted additional experiments on smaller language models. While our main experiments employed LLMs like Qwen3-32B~\citep{qwen2025qwen25technicalreport} and Gemini 2.0 Flash~\citep{google2025gemini}, it remains a question whether our method generalizes effectively to more compact language models in the 7B–8B parameter range. We test two smaller models, Mistral-7B~\citep{jiang2023mistral7b} and Llama-3.1-8B-Instruct~\citep{grattafiori2024llama} as reranking backbones and the results are shown in Table~\ref{tab:backbone}.

\begin{table}[!ht]
\centering

\caption{Reranking Performance with different small language models as backbones on LitSearch.}

\begin{tabular}{llcc}
\toprule[1pt]
\textbf{Reranking Method} & \textbf{Model Backbone} & 
\textbf{N@10} & 
\textbf{R@10} \\
\midrule[0.7pt]
None & None & 14.2 & 21.2 \\
\midrule[0.7pt]
Vanilla & Mistral-7B & 20.9 & 24.6 \\
\Method & Mistral-7B & 24.8 & 28.4 \\
\midrule[0.7pt]
Vanilla & Llama-3.1-8B & 22.1 & 25.3 \\
\Method & Llama-3.1-8B & 26.0 & 31.2 \\
\midrule[0.7pt]
Vanilla & RankMistral & 20.1 & 24.3 \\
Vanilla & RankVicuna & 26.5 & 32.4 \\
Vanilla & RankZephyr & 34.1 & 39.0 \\
\bottomrule[1pt]
\end{tabular}

\label{tab:backbone}
\vspace{-1.5em}
\end{table}

Our experiments confirmed two main findings. First, \Method \ remained effective on smaller language models, consistently outperforming initial retrieval and vanilla reranking baseline. Second, although it did not surpass the supervised models in absolute performance, this outcome was expected, since the supervised methods require extensive training resources. For example, RankVicuna~\citep{pradeep2023rankvicunazeroshotlistwisedocument} and RankZephyr~\citep{pradeep2023rankzephyreffectiverobustzeroshot} use GPT-3.5-distilled training data (1,000 K MS MARCO instances) and significant computational resources (320 GPU-hours on NVIDIA RTX A6000s). \Method \ holds key practical advantages: it requires no additional training data or computational resources, works seamlessly as a plug-and-play method across different models and domains, and ensures data privacy. These strengths underscore \Method's value despite its lower absolute scores compared to supervised models.

\subsubsection{Human Evaluation on Feature Quality}
\label{human evaluation}

To more directly assess the quality of our zero-shot LLM-based information extraction—beyond the indirect evidence from retrieval performance and case in Figure~\ref{fig:representation_pipe}, we conducted two human evaluation studies: (i) Human Verification and (ii) Human–LLM Comparison. 

\para{Human Verification.} Three domain experts independently evaluated the correctness of LLM-extracted categories, topics, sections, and keywords for a held-out set of 29 documents. For categories and topics, we measured accuracy, defined as the proportion of LLM-assigned labels that matched the human-annotated ground truth. For sections and keywords, which can have multiple correct instances per document, we assessed two complementary aspects: (i) Faithfulness – defined as 1 – hallucination rate, where hallucination rate is the proportion of elements generated by the LLM that do not actually appear in the source document; (ii) Completeness – defined as 1 – missing rate, where missing rate is the proportion of ground-truth elements in the document that the LLM failed to identify. This evaluation covered 29 categories, 29 topics, 197 section instances, and 623 keyword instances. All assessments were performed using the full text of each document and without exposure to other annotators’ judgments.

\para{Human–LLM Comparison.}
To measure subjective preference, annotators were presented with unlabeled pairs of annotations—one produced by a human annotator and one by the LLM—for 15 randomly selected documents. The identity of each annotation source was concealed, and annotators indicated which version they preferred for overall quality and informativeness. The primary metric was win rate, defined as the fraction of instances in which the LLM’s output was preferred over the human’s.

\para{Findings.}
As exhibited in Table~\ref{human evaluation}, the LLM achieved high accuracy for category and topic assignments, strong faithfulness, and high completeness for both sections and keywords. The win rate in the comparison study indicates that, even without explicit optimization, the LLM’s outputs are competitive with human annotations. Overall, these results demonstrate that our zero-shot IE pipeline produces high-quality structured information across multiple content dimensions, providing direct evidence that the extraction process is not a limiting factor for our retrieval improvements.

% ...

\section{Conclusion}
In this work, we first identify fundamental challenges of standard listwise reranking in scientific retrieval. On the one hand, the first-stage retriever often fails to rank truly relevant documents high due to the lack of domain-specific training data. On the other hand, existing LLM-based listwise reranking methods typically operate over full-document inputs, which are substantial in token usage and therefore limited to a small number of candidates. As a result, when the first-stage retrieval is suboptimal for scientific retrieval, the performance of standard reranking method is restricted.

To address this, we propose \Method, a training-free and model-agnostic reranking framework for scientific retrieval. \Method~employs compact semantic features for coarse-grained reranking. This allows the LLM to consider a broader range of candidates within its context window. A subsequent fine-grained reranking stage then refines the top results using full text inputs. Experiments on five scientific datasets show that \Method~significantly improves the reranking performance.

Overall, we explore semantic features as compact document representations for LLM reranking in scientific retrieval. Our results demonstrate how integrating semantic feature extraction into the retrieval pipeline creates a synergistic interaction between information extraction and reranking, leading to more effective scientific literature retrieval.

\section*{Ethical Impact Statement}
This work focuses on improving scientific document retrieval using publicly available datasets, and we do not foresee direct harm to individuals or groups. Nonetheless, LLM-based reranking methods can inherit biases from underlying corpora, potentially overrepresenting certain disciplines or communities. They could also, in real-world deployments, surface sensitive or dual-use technical content. To mitigate these risks, we recommend applying fairness audits, integrating content filters for high-risk topics, and maintaining human oversight in the interpretation of retrieved results. Additionally, caution should be taken to avoid overreliance on automated rankings, ensuring they complement rather than replace critical human judgment.



\balance

\bibliographystyle{ACM-Reference-Format}
\bibliography{references} 


\end{document}